\documentclass[a4paper]{jpconf}
\usepackage{graphicx}
\usepackage{citesort}
\usepackage[square,sort&compress,numbers]{natbib}
\usepackage[braket,qm]{qcircuit}
\usepackage{amsmath}
\usepackage[rightcaption]{sidecap}
\usepackage[colorlinks=true,citecolor=blue]{hyperref}

\begin{document}
\title{Particle track reconstruction with noisy intermediate-scale quantum computers}

\author{T Schw\"agerl$^{1,2}$, C Issever$^{1,2}$, K Jansen$^2$, T J Khoo$^1$, S K\"uhn$^2$, C T\"uys\"uz$^{1,2}$ and H Weber$^1$}

\address{$^1$ Institut f\"ur Physik, Humboldt-Universit\"at zu Berlin, Newtonstr. 15, 12489 Berlin, Germany}
\address{$^2$ Deutsches Elektronen-Synchrotron DESY, Platanenallee 6, 15738 Zeuthen, Germany}

\ead{tim.schwaegerl@physik.hu-berlin.de}

\begin{abstract}
The reconstruction of trajectories of charged particles is a key computational challenge for current and future collider experiments. Considering the rapid progress in quantum computing, it is crucial to explore its potential for this and other problems in high-energy physics. The problem can be formulated as a quadratic unconstrained binary optimization (QUBO) and solved using the variational quantum eigensolver (VQE) algorithm. In this work the effects of dividing the QUBO into smaller sub-QUBOs that fit on the hardware available currently or in the near term are assessed. Then, the performance of the VQE on small sub-QUBOs is studied in an ideal simulation, using a noise model mimicking a quantum device and on IBM quantum computers. This work serves as a proof of principle that the VQE could be used for particle tracking and investigates modifications of the VQE to make it more suitable for combinatorial optimization.
\end{abstract}

\section{Introduction}
\begin{SCfigure}[0.75][ht]
    \caption{Hits and reconstructed trajectories of particles in the transverse plane of the detector. The detector layers are indicated by the grey concentric circles. Real tracks (indicated in green) are successfully reconstructed. Tracking algorithms can miss or make up tracks, which are indicated in blue and red. Generated using the Python framework hepqpr-qallse~\cite{Linder2023}.}
    \label{fig:event_ds10}
    \includegraphics[width=0.48\textwidth]{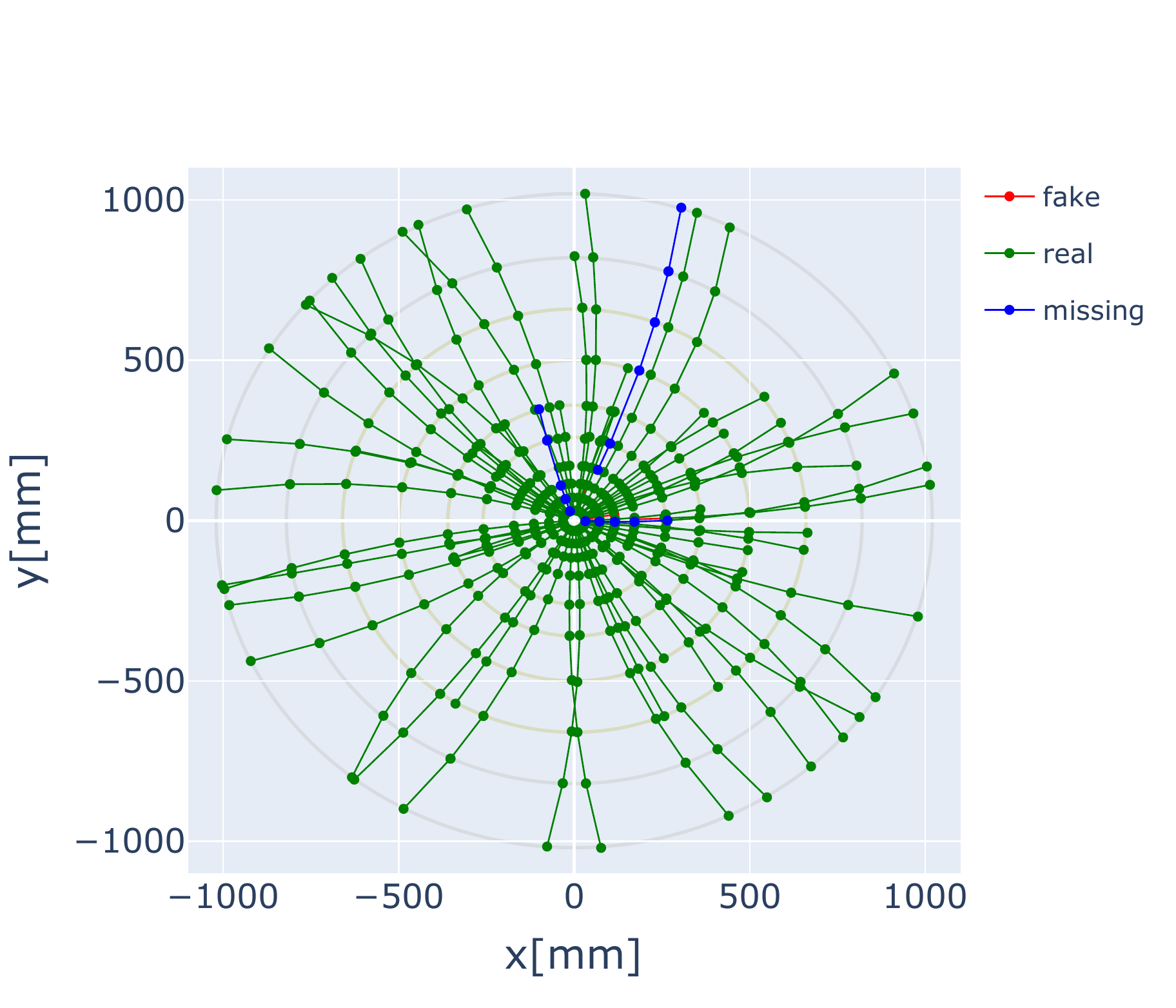}
\end{SCfigure}

At the Large Hadron Collider (LHC) thousands of particles are produced in collisions of two opposing proton beams. The collision points are enclosed by detectors whose inner-most components are multiple silicon layers with cylindrical symmetry around the beam-pipe. One key computational challenge is particle tracking, i.e.\ the reconstruction of trajectories of charged particles from small energy deposits in the detector, referred to as hits. In Fig.~\ref{fig:event_ds10} hits and reconstructed trajectories of particles in the transverse $x-y$-plane of the detector are shown, perpendicular to the beam-direction.
Future colliders, such as the high-luminosity LHC (HL-LHC)~\cite{Apollinari2015}, are in general designed to produce many particles, rendering the reconstruction more difficult and asking for improvements, especially in terms of CPU resources. 
Quantum computing is an aspiring technology enabling exponential speedups for certain problems~\cite{Shor1997}. Currently available quantum devices can have up to several hundred qubits, but still suffer from a considerable level of noise, and are commonly referred to as noisy intermediate-scale quantum (NISQ) hardware~\cite{Preskill2018}. One algorithm for NISQ computers is the hybrid quantum-classical variational quantum eigensolver (VQE) that was originally proposed to estimate the ground state energy of molecules~\cite{Peruzzo2014}. In our study, the potential of VQE to solve the problem of particle track reconstruction is investigated. There are several works on quantum algorithms for particle tracking that investigate different approaches~\cite{Bapst2019, Tueysuez2021, Duckett2022, Funcke2023, Zlokapa2021}. For a review of quantum computing for data analysis in high-energy physics, we refer to Ref.~\cite{Delgado2022a}. In Ref.~\cite{Bapst2019}, the problem was formulated as a quadratic unconstrained binary optimization (QUBO) problem, tailored for D-Wave quantum annealers. In this work, we use the same formulation and investigate the performance of the VQE on gate-based quantum computers to solve the QUBO.

The paper is organized as follows. In Sec.~\ref{sec:methodology}, we explain the QUBO formulation of particle tracking, our approach for dividing the full QUBO into smaller sub-QUBOs, and the modifications of the VQE for combinatorial problems we use. In Sec.~\ref{sec:results}, we study the impact of sub-QUBO size on the efficiency and the purity, and show results for the performance of VQE on sub-QUBO level in an ideal simulation, using a noise model mimicking a quantum device and on real quantum hardware. Finally, we conclude in Sec.~\ref{sec:conclusion}.

\section{Methodology\label{sec:methodology}}
This work builds upon the approach introduced in Ref.~\cite{Bapst2019} using the data set of the TrackML challenge \cite{Kaggle}. We use its implementation in the Python framework hepqpr-qallse~\cite{Linder2023} to construct the QUBO and to compute efficiency and purity. The main idea is to reconstruct the trajectories from smaller track segments. The smallest constituents of tracks are single hits, but it is difficult to reconstruct tracks from hits directly. Two hits $a$ and $b$ in different layers of the detector form a doublet $(a, b)$. Doublets $(a, b)$ and $(b, c)$ are combined to triplets $(a, b, c)$. Now, the algorithm aims at identifying true $(t)$ triplets that are part of trajectories of charged particles in contrast to false $(f)$ triplets that are random combinations of three hits. This is done by rating the quality of individual triplets and by comparing triplets to each other. One key component is to identify combinations of triplets $(a, b, c)$ and $(b, c, d$) that overlap by two hits and form a quadruplet $(a, b, c, d)$.

\subsection{Quadratic unconstrained binary optimization\label{sec:qubo}}
QUBO problems are combinatorial optimization problems arising in a wide range of applications. In particular, the problem of particle tracking can be expressed as a QUBO with a cost function $Q(T)$, where $T$ is a binary vector of length $N$ whose entries correspond to the triplet candidates $T_i$. Triplets identified by the algorithm to be true (false) triplets are called positives $(p)$ (negatives $(n)$) and are denoted $T_i=1$ ($T_i=0$). Triplet parameters, such as the curvature, are used to compute the coefficients $a_i$ that rate the quality of individual triplets, and the coefficients $b_{ij}$ that express the compatibility of two triplets. The details of the computations can be found in Ref.~\cite{Bapst2019}. 
The coefficients are constructed such that the minimum of the function $Q(T)$ holds preferably all true  and no false triplets. Using these coefficients, particle tracking is reduced to the problem of finding the minimum of the cost function
\begin{align}
    Q(T)=\sum_i^N a_i T_i + \sum_i^N\sum_{j<i}^N b_{ij}T_i T_j.
    \label{eq:cost_function}
\end{align}
Due to the limited size of current quantum devices, we use a geometric approach to divide the full QUBO into smaller sub-QUBOs that fit on NISQ computers. The triplets are sorted according to their angle in the $x-y$-plane, and slices of $k$ consecutive triplets are selected. The slices overlap by $k/2$ triplets. Subsequently, sub-QUBOs are created for each slice. We chose this approach to give a perspective on how large the sub-QUBOs have to be to reasonably solve the tracking problem. $Q(T)$ can be transformed to a Hamiltonian by mapping $T_i\to (1-Z_i)/2$, where $Z_i$ denotes the third Pauli matrix. Every triplet candidate is represented by one qubit. Now, VQE can be used to find the ground state of the Hamiltonian and thus the minimum of $Q(T)$. 

\subsection{Layer variational quantum eigensolver}
\label{sec:lvqe}
\begin{figure*}
    \centering
    \begin{align*}
        \Qcircuit @C=0.25em @R=.1em {
        \lstick{\ket{0}} & \gate{R_Y(\vartheta_1)} & \qw & \ctrl{1} & \gate{R_Y(\vartheta_5)} & \ctrl{1} & \gate{R_Y(\vartheta_9)} & \qw & \qw & \qw & \qw & \qw & \quad & \quad & \quad & \ctrl{1} & \qw & \qw & \gate{R_Y(\vartheta_5)} & \qw & \qw & \ctrl{1} & \qw\\
        \lstick{\ket{0}} & \gate{R_Y(\vartheta_2)} & \qw & \targ & \gate{R_Y(\vartheta_6)} & \targ & \gate{R_Y(\vartheta_{10})} & \ctrl{1} & \gate{R_Y(\vartheta_{13})} & \ctrl{1} & \gate{R_Y(\vartheta_{15})} & \qw & \quad & \quad & \quad & \targ & \ctrl{1} & \qw & \gate{R_Y(\vartheta_6)} & \qw & \ctrl{1} & \targ & \qw\\ 
        \lstick{\ket{0}} & \gate{R_Y(\vartheta_3)} & \qw & \ctrl{1} & \gate{R_Y(\vartheta_7)} & \ctrl{1} & \gate{R_Y(\vartheta_{11})} & \targ & \gate{R_Y(\vartheta_{14})} & \targ & \gate{R_Y(\vartheta_{16})} & \qw & \quad & \quad & \quad & \qw & \targ & \ctrl{1} & \gate{R_Y(\vartheta_7)} & \ctrl{1} & \targ & \qw & \qw\\
        \lstick{\ket{0}} & \gate{R_Y(\vartheta_4)} & \qw & \targ & \gate{R_Y(\vartheta_8)} & \targ & \gate{R_Y(\vartheta_{12})} & \qw & \qw & \qw & \qw & \qw & \quad & \quad & \quad & \qw & \qw & \targ & \gate{R_Y(\vartheta_8)} & \targ & \qw & \qw & \qw
        \gategroup{1}{2}{4}{2}{.9em}{--} 
        \gategroup{1}{2}{4}{12}{.9em}{--}
        \put(-300, -50){(a)}
        \put(-75, -50){(b)}
        }
    \end{align*}
    \caption{The circuit of the L-VQE algorithm (a). The circuit consists of single-qubit $R_Y$ rotation gates and entangling two-qubit CNOT gates. The ansatz is grown by adding layers (dashed box on the right hand side) to the initial rotation layer (dashed box on left hand side). Layer with fewer $R_Y$ gates used for the runs on real hardware (b). Figure adapted from Ref.~\cite{Liu2022}.}
    \label{fig:circuit}
\end{figure*}
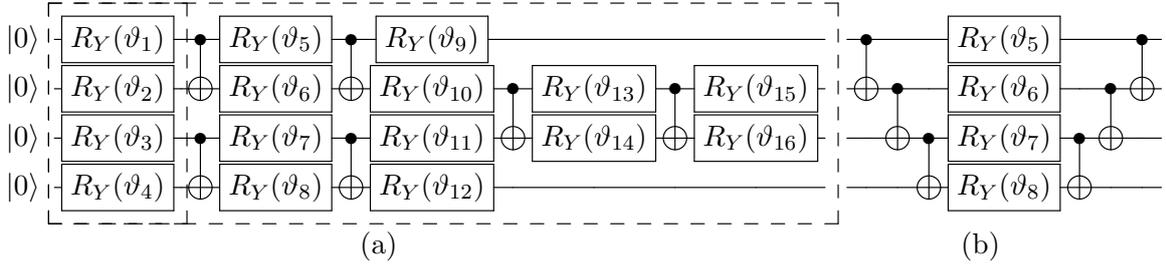

In Ref.~\cite{Liu2022} the Layer VQE (L-VQE) approach was introduced for combinatorial optimization problems.
The circuit structure for the L-VQE approach is shown in Fig.~\ref{fig:circuit}.
First, a simple rotation layer of $R_Y$ gates is initialized with random parameters uniformly drawn from the interval $[0, 2\pi]$. Then, the VQE is run for this single layer. The optimization process is stopped after a fixed number of iterations, and not carried out until full convergence. A second layer including $R_Y$ gates and CNOT gates is added to the circuit. Its initial parameters are set to zero, not affecting the state reached by the previous optimization. The optimization is then carried out for all parameters, again for a fixed number of iterations. This procedure is repeated for every additional layer, and in the last step optimization is carried out until convergence. In principle, the initial rotation layer is sufficient to express the solution, since it is a computational basis state because the Hamiltonian is diagonal. However, the optimization for such an ansatz gets easily stuck in local minima~\cite{Anschuetz2022}. The idea behind the additional layers is, that they might allow the algorithm to escape from these minima.

\subsection{Conditional value at risk cost function}
\label{sec:cvar}
Originally, the VQE was designed to approximate the ground state of a Hamiltonian by minimizing the energy expectation value. For diagonal Hamiltonians it is enough to reach a state with sufficiently large ground state component that can be revealed by taking enough measurements. Assuming the final VQE solution has 1\% ground state component and we take 1024 measurements, the probability for obtaining the optimal solution in one of the measurements is greater than 99.996\%. Thus, the authors of Ref.~\cite{Barkoutsos2020} proposed to use a conditional value at risk (CVaR) as a cost function, which uses the fraction $\alpha$ of measurements with the lowest energy instead of the full sample mean.

\section{Results\label{sec:results}}
The geometric sub-QUBO approach allows for rating the full tracking algorithm's performance and the performance of VQE on sub-QUBO level separately.
This is an important step because current quantum computers are still of limited size, thus rendering it unpractical to run the full tracking algorithm on hardware devices. First, we introduce the relevant metrics to rate the tracking performance and the performance of VQE on sub-QUBO level. 
Subsequently, we present results for the tracking performance for different sub-QUBO sizes and for different numbers of particles per event, ranging from current conditions at the LHC (around 2k particles per event) to conditions expected at the HL-LHC. Finally, we show results of VQE for small sub-QUBOs and assess how the previously introduced modifications of VQE affect its performance.

\subsection{Performance metrics}
To rate the tracking algorithm's performance and to investigate the impact of dividing the full QUBO into smaller sub-QUBOs, we compute the $\text{efficiency} = \frac{tp}{tp+fn}$ and the $\text{purity} = \frac{tp}{tp+fp}$ defined by the true positives $(tp)$, the false positives $(fp)$ and the false negatives $(fn)$ in the final set of doublets. For details on which doublets that are included in the final set of triplets are considered in the rating of the algorithm's performance, we refer to Ref.~\cite{Bapst2019}. 
To rate the performance of VQE on sub-QUBO level, we calculate the fraction of instances with at least 1\% ground state component.

\subsection{Impact of the sub-QUBO size}
\begin{figure}
    \centering
    \includegraphics[width=0.48\textwidth]{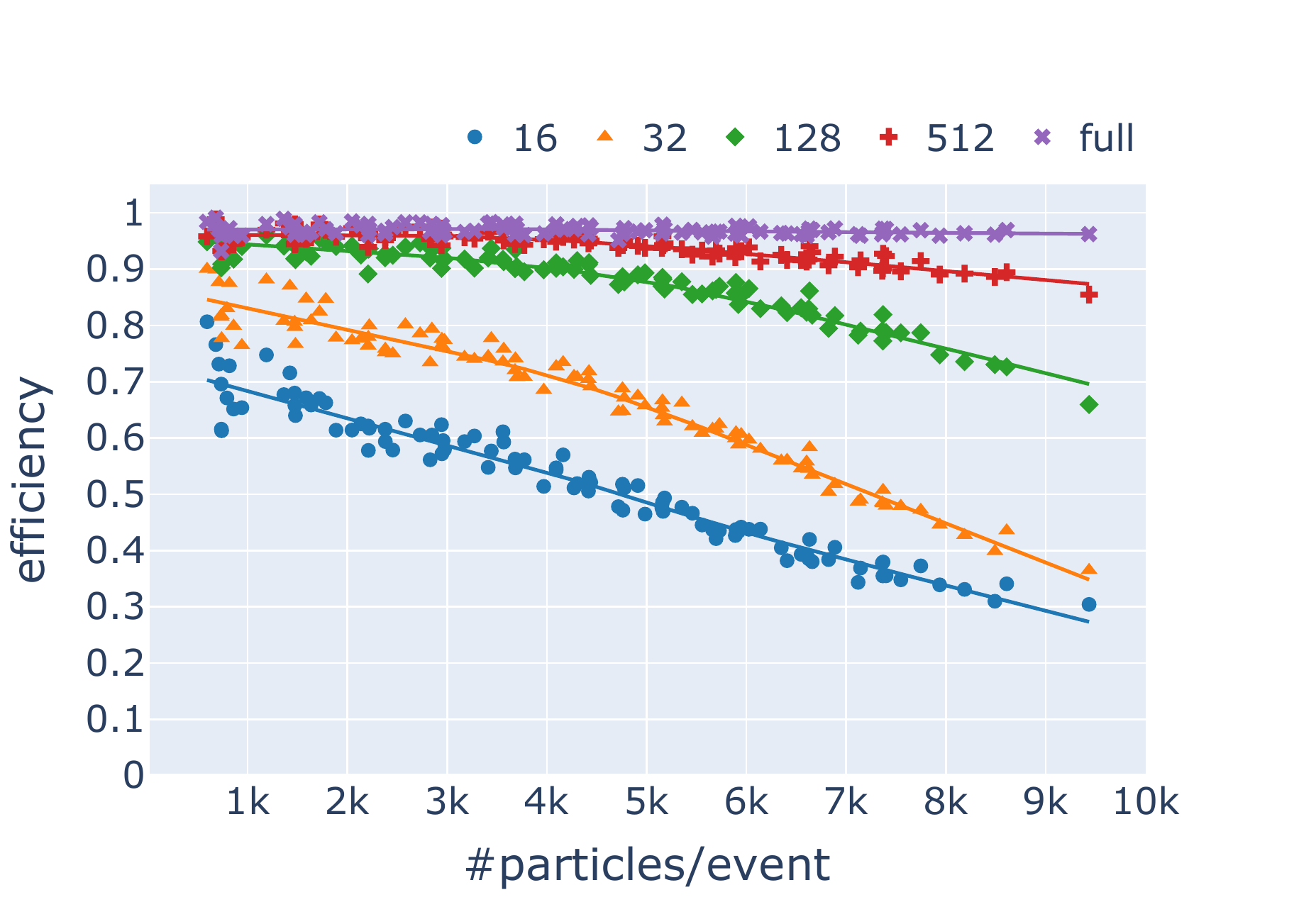}
    \quad
    \includegraphics[width=0.48\textwidth]{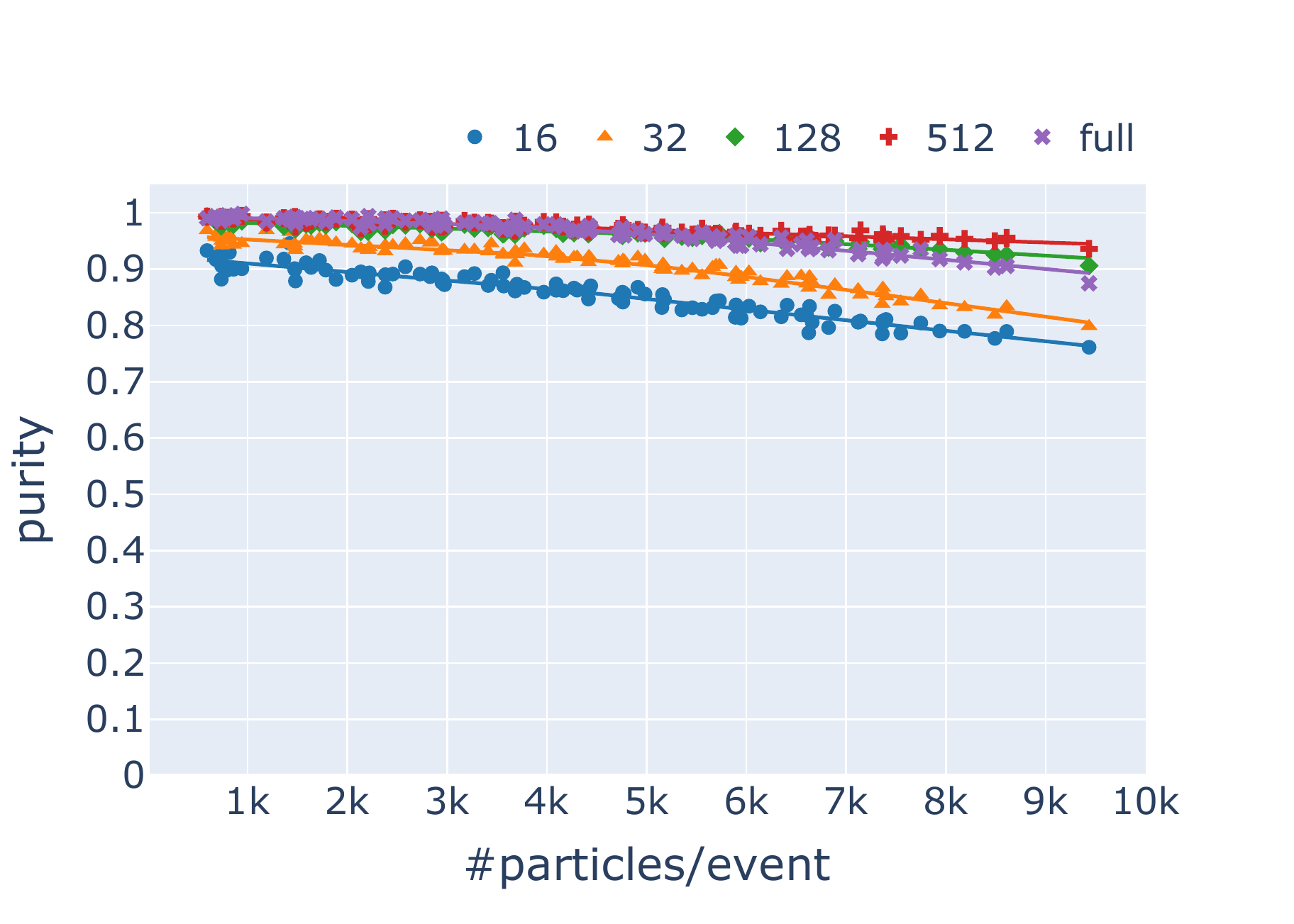}
    \put(-455,120){(a)}
    \put(-220,120){(b)}
    \caption{Efficiency (a) and purity (b) as a function of track density for slices of increasing size in the $x-y$-plane and for the full QUBO. For slices of size 512 and small track densities, the performance is comparable to solving the full QUBO.}
    \label{fig:precision_recall}
\end{figure}

To assess the impact of the sub-QUBO size on the algorithm's performance, various sub-QUBO sizes are studied and solved using neal~\cite{Neal2023}, an implementation of a simulated annealing sampler.
All triplets that are identified as positives in at least one of the overlapping slices are marked as positives. Then, efficiency and purity are computed. This estimates the performance of VQE for particle tracking under the assumption that VQE finds the ground state of every sub-QUBO. Efficiency and purity for sub-QUBO sizes ranging from 16 to 512 triplets as well as for the full QUBO are shown in Fig.~\ref{fig:precision_recall}.
For sub-QUBO size 16 the efficiency reaches 0.7 for the lowest track densities and then drops rapidly. Purity is above 0.8 for almost the full range of track densities. Doubling the size to 32 improves efficiency and purity. The performance of the VQE for slices of size 16 to 32 is examined in Sec.~\ref{sec:vqeperformance}. For size 128, the  efficiency is above 0.9 for up to 5k particles per event and purity is comparable to solving the full QUBO. A sub-QUBO size of 512 is comparable to solving the full QUBO for up to 5k particles per event in terms of efficiency and purity. 

\subsection{Performance of the VQE on sub-QUBO level}
\label{sec:vqeperformance}
\begin{SCfigure}[0.9][ht]
    \caption{Fraction of instances with at least 1\% ground state component at the end of the VQE as a function of sub-QUBO size in an ideal simulation, using the full noise model of the quantum computer ibmq\_kolkata and the real device. Results for two values of $\alpha$ in the CVaR cost function and for up to two additional layers in L-VQE are shown. Sizes 28 and 32 show only the ideal scenario.}
    \label{fig:fraction}
    \includegraphics[width=0.65\textwidth]{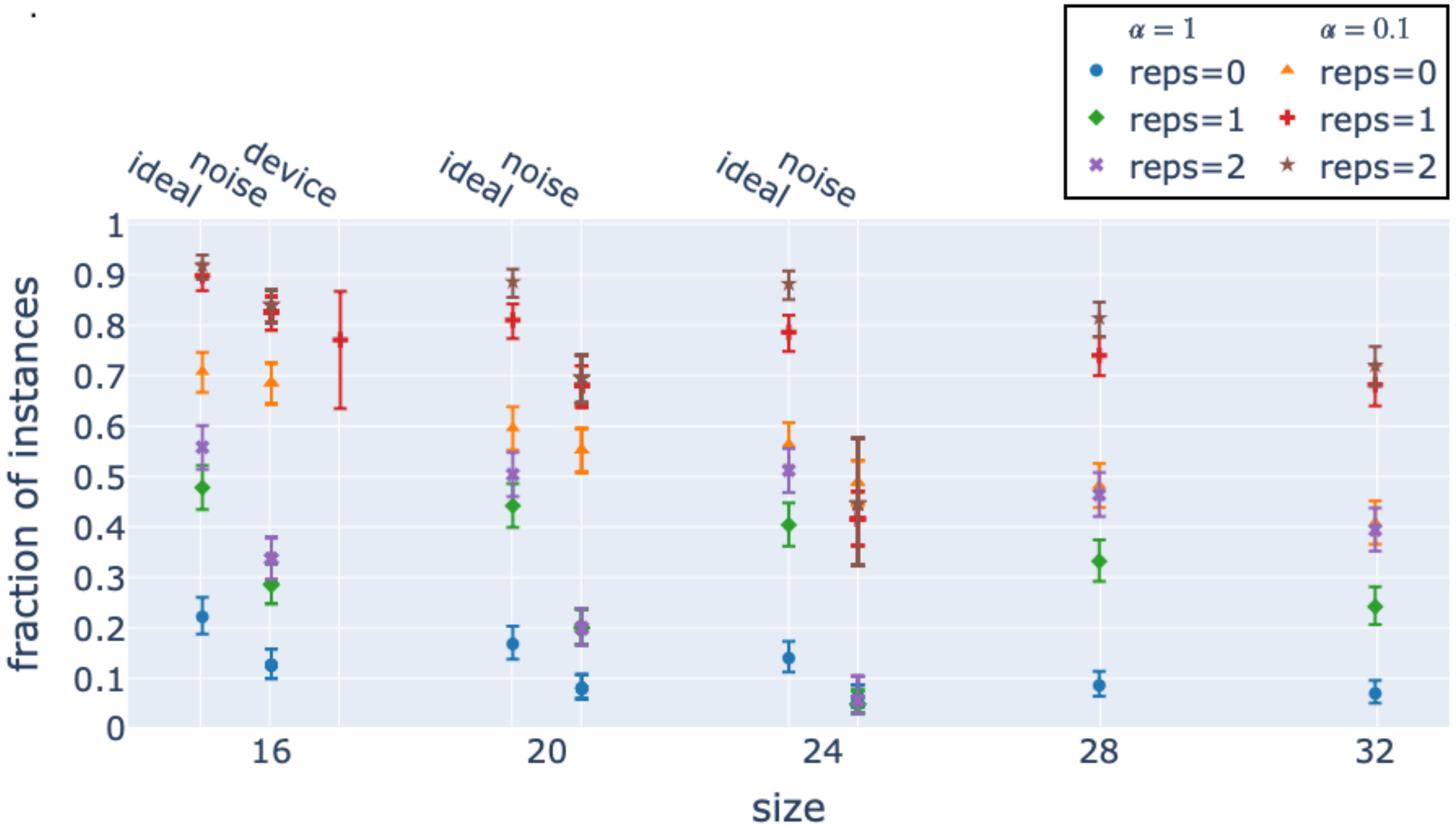}\,
\end{SCfigure}

The performance of the VQE for small sub-QUBOs of size 16 to 32 and for track densities of 1k particles per event is determined for three scenarios: i) an ideal simulation assuming a perfect, noise free quantum computer, ii) using a noise model mimicking the quantum computer ibmq\_kolkata, and iii) on the real device. We use Qiskit~\cite{Qiskit} for the simulations and hardware runs.  Moreover, we adapt the L-VQE approach with up to 2 additional layers and the CVaR cost function with $\alpha=0.1$ and $\alpha=1$ (the latter corresponding to running a conventional VQE). The number of reps indicates the additional layers that are added to the initial rotation layer. For the initial rotation layer and after adding an additional layer, 128 iterations of the VQE are performed. Then, the circuits are further optimized until a total of 1024 iterations is reached. In every iteration we use 1024 measurements to evaluate the cost function in all three scenarios. To update the circuit parameters we use Constrained Optimization BY Linear Approximation (COBYLA)~\cite{Powell1998}. In Fig.~\ref{fig:fraction} the mean and 95\% confidence interval of fractions of instances with at least 1\% ground state component for 50 slices (5 slices on real hardware) and 10 random initial points are shown.
Using the CVaR as a cost function with $\alpha=0.1$ yields a significant improvement for the ideal and noisy simulations. In the ideal setting, additional layers are beneficial for all system sizes. In the noisy simulation adding one additional layer yields an improvement for up to 20 qubits. For 32 qubits, around 70\% of instances reach at least 1\% ground-state component in an ideal simulation. The data obtained on the real quantum device is compatible with our classical simulation.

\section{Discussion and outlook}
\label{sec:conclusion}
This work serves as a proof of principle that the VQE can be used for particle tracking. 
It was shown that sub-QUBO sizes compatible with NISQ hardware yield reasonable efficiency and purity.  
In an ideal simulation without noise, the VQE finds the solution of small sub-QUBOs for the majority of instances. 
Noise, either coming from a noise model mimicking a real quantum device or the quantum hardware, significantly affects the results. Thus, in future studies we aim at investigating different techniques for error mitigation.
Two modifications of VQE, the CVaR cost function and L-VQE, were tested. CVaR shows a clear advantage for combinatorial optimization. While the additional layers in L-VQE also show a small advantage, the role of entanglement for classical optimization problems remains an open question asking for further research.

\ack
This work is supported by: 
i) the Einstein Research Unit -``Perspectives of a quantum digital transformation: Near-term quantum computational devices and quantum processors'', 
ii) the European Union’s Horizon Europe research and innovation funding programme under the ERA Chair scheme with grant agreement No. 101087126 and under the European Research Council (ERC) with grant agreement No. 787331,
iii) the Helmholtz Association -``Innopool Project Variational Quantum Computer Simulations (VQCS)'',
and iv) the Ministry of Science, Research and Culture of the State of Brandenburg within the Centre for Quantum Technologies and Applications (CQTA). \raisebox{-1.0em}{\includegraphics[width = 0.05\textwidth]{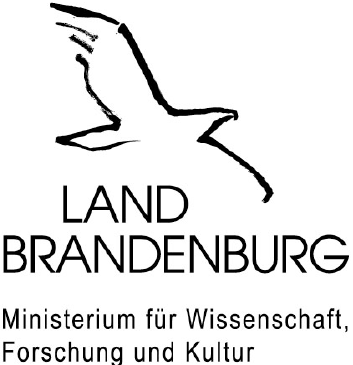}}

\bibliography{iopart-num}

\end{document}